\documentclass[aps, twocolumn, letterpaper, superscriptaddress, nofootinbib, pra]{revtex4-1}

\usepackage{amsmath}
\usepackage{amssymb}
\usepackage{mathrsfs}
\usepackage{xspace}
\usepackage{braket}
\usepackage{bbold}
\usepackage{nicefrac}
\usepackage{color}

\usepackage{ifpdf}
\ifpdf
\fi

\newcommand{\eq}[1]{(\ref{#1})}
\newcommand{\Eq}[1]{Eq.~(\ref{#1})}
\newcommand{\Eqs}[1]{Eqs.~(\ref{#1})}

\newcommand{\Sec}[1]{Sec.~\ref{#1}}
\newcommand{\myRef}[1]{Ref.~\cite{#1}}
\newcommand{\Refs}[1]{Refs.~\cite{#1}}

\newcommand{\eg}{{e.g.,\/}\xspace}
\newcommand{\ie}{{i.e.,\/}\xspace}
\newcommand{\etal}{{\it et~al.\/}\xspace}

\newcommand{\pd}{\partial}
\newcommand{\del}{\nabla}
\newcommand{\mc}[1]{\mathcal{#1}}
\newcommand{\mcc}[1]{\mathfrak{#1}}
\newcommand{\msf}[1]{\mathsf{#1}}
\newcommand{\mcu}[1]{\mathscr{#1}}
\newcommand{\oper}[1]{\smash{\widehat{#1}}}
\newcommand{\boper}[1]{\oper{\boldsymbol{#1}}}
\renewcommand{\vec}[1]{{\boldsymbol{#1}}}
\newcommand{\favr}[1]{\langle #1 \rangle}
\newcommand{\hm}{a}

\newcommand{\ee}{\mathrm{e}}
\newcommand{\ii}{\mathrm{i}}
\newcommand{\dd}{\mathrm{d}}

\renewcommand{\Re}{\text{Re}\,}

\newcommand{\sk}{\mathsf{k}}
\newcommand{\tS}{S}

\newcommand{\lie}{\mathrm{\text{\pounds}}}
\newcommand{\total}[1]{\msf{#1}}

\definecolor{darkred}{rgb}{0.75, 0, 0}
\usepackage[colorlinks=true, linkcolor=darkred, citecolor=darkred, hypertexnames=false]{hyperref}

\begin{document}

\title{Gauge-invariant gravitational waves in matter beyond linearized gravity}

\author{Deepen Garg}
\affiliation{Department of Astrophysical Sciences, Princeton University, Princeton, New Jersey 08544, USA}
\author{I.\ Y.\ Dodin}
\affiliation{Department of Astrophysical Sciences, Princeton University, Princeton, New Jersey 08544, USA}
\affiliation{Princeton Plasma Physics Laboratory, Princeton, NJ 08543, USA}

\date{\today}

\begin{abstract}
Modeling the propagation of gravitational waves (GWs) in media other than vacuum is complicated by the gauge freedom of linearized gravity in that, once nonlinearities are taken into consideration, gauge artifacts can cause spurious acceleration of the matter. To eliminate these artifacts, we propose how to keep the theory of dispersive GWs gauge-invariant beyond the linear approximation and, in particular, obtain an unambiguous gauge-invariant expression for the energy--momentum of a GW in dispersive medium. Using analytic tools from plasma physics, we propose an exactly gauge-invariant ``quasilinear'' theory, in which GWs are governed by linear equations and also affect the background metric on scales large compared to their wavelength. As a corollary, the gauge-invariant geometrical optics of linear dispersive GWs in a general background is formulated. As an example, we show how the well-known properties of vacuum GWs are naturally and concisely yielded by our theory in a manifestly gauge-invariant form. We also show how the gauge invariance can be maintained within a given accuracy to an arbitrary order in the GW amplitude. These results are intended to form a physically meaningful framework for studying dispersive GWs in matter.
\end{abstract}

\maketitle

\section{Introduction}
\label{sec:intro}

Analytical studies of gravitational waves (GWs) usually focus on the nondispersive vacuum modes propagating at the speed of light \cite{ref:flanagan05}, which are also the relevant modes for the observations so far \cite{ref:abbott17e, ref:abbott19b, ref:abbott19c}. However, in the early Universe and possibly also near compact objects, the coupling of GWs with matter can be nonnegligible. Understanding this coupling is potentially important, for example, for understanding the electromagnetic signatures of the GW radiation \cite{ref:gieg19, ref:adshead21}. Also, since this coupling is small for the tensor modes and waves in cold matter \cite{ref:flauger18}, even small corrections due to thermal effects \cite{ref:kumar19}, viscosity \cite{ref:madore73}, and matter-induced modification of the wave polarization \cite{ref:moretti20, my:gwjeans} may be significant.

The tools for modeling dispersive GWs in matter can be imported from electrodynamics, plasma physics in particular, where the wave-induced oscillations of matter are commonly described in terms of electric susceptibility \cite{book:stix}. Provided that matter is continuously distributed in spacetime, its effect on a GW can be similarly, and conveniently, described in terms of gravitational susceptibility \cite{my:gwjeans, my:gwponder}. Then, by analogy with electrodynamics of continuous media \cite{book:landau8}, two questions arise: (i) How can one derive the equations of GW propagation at nonzero gravitational susceptibility of the ambient medium? (ii) How can one describe the exchange of the energy--momentum between dispersive GWs and matter?

For small wave amplitudes, the first question can be answered within linear theory \cite{my:gwjeans}, but the second question cannot. The energy--momentum of a linear wave produces nonlinear average forces on the underlying medium, and is itself quadratic in the wave amplitude. Defining it generally requires so-called \textit{quasilinear} (QL) theory \cite{ref:dewar73, ref:kaufman87b, my:ql}, specifically, QL theory of adiabatic (nonresonant) wave--medium interactions. Such interactions are commonly described using variational methods \cite{ref:cary81} that originate from Whitham's average-Lagrangian approach \cite{ref:whitham65, book:whitham}. Specifically, an adiabatic response of a medium to a wave is expressed through the wave field and the medium susceptibility; then the wave Lagrangian is identified as the part of the wave--matter Lagrangian that is quadratic in the wave amplitude, whence the wave energy--momentum and the nonlinear forces on the medium are readily inferred in the usual manner \cite{my:amc}. This approach has been particularly fruitful in plasma theory in application to the electromagnetic interactions  \cite{ref:dougherty70, ref:dewar70, ref:dewar77, my:qponder, my:itervar, phd:ruiz17}, where the wave energy--momentum and the nonlinear forces on the medium are naturally expressed through the electric field  \cite{my:amc}. Whitham's approach has also been used to describe GWs \cite{ref:isaacson68a, ref:maccallum73, ref:araujo89, ref:butcher09, ref:andersson21} and allows to circumvent the various problems \cite{ref:green11, tex:green15, ref:buchert15, ref:kaspar12, ref:clarkson11, phd:kaspar14} associated with averaging tensor fields in a self-consistent covariant fashion \cite{ref:zalaletdinov96}. Still, application of this approach to GWs remains deficient in that, unlike for electromagnetic waves, GW's average Lagrangians generally lack gauge invariance \cite{ref:isaacson68a}.

As a reminder, the gauge invariance for GWs derives from general covariance \cite[Sec.~7.1]{book:carroll}. Consider a background metric $\smash{g_{\alpha\beta}} = \mc{O}(1)$ and a perturbation metric $\smash{h_{\alpha\beta}} = \mc{O}(a)$ on top of it, where $a \ll 1$ is a small parameter. Then, a coordinate transformation $\smash{x^\mu \to x'^\mu = x^\mu + \xi^\mu}$, with $\smash{\xi^\mu} = \mc{O}(a)$, implies $\smash{g_{\alpha\beta}} \to \smash{g'_{\alpha\beta}} = \smash{g_{\alpha\beta}}$ and $\smash{h_{\alpha\beta}} \to \smash{h'_{\alpha\beta}} = \smash{h_{\alpha\beta}} - \smash{\lie_\xi g_{\alpha\beta}}$, where $\smash{\lie_\xi}$ is the Lie derivative along the vector field within linearized theory (\Sec{sec:ct}). If one considers $\smash{h_{\alpha\beta}}$ as a tensor field on the background spacetime $\smash{g_{\alpha\beta}}$, this transformation can be viewed as a gauge transformation (with $\xi^\mu$ being the gauge field) and, by general covariance, cannot have measurable effects. However, the GW Lagrangian is normally formulated in terms of $\smash{h^{\alpha\beta}}$ \cite{ref:butcher09} and not gauge-invariant in the presence of matter \cite{ref:isaacson68a}. That leads to a non-invariant energy--momentum of the wave and the possibility of spurious acceleration (or heating) of the matter with gauge artifacts, which is unacceptable. This brings the question: how can one construct a gauge-invariant QL theory of GWs and, in particular, define the energy--momentum of a dispersive GW such that it would be \textit{exactly} free of gauge artifacts?

Here, we show how to do this using the tools that we have developed previously in \Refs{my:gwjeans, my:ql, my:gwponder, my:gwinvar, tex:mygwplasma}. Let us briefly describe this series of papers to put our work into perspective. Our general formalism for studying dispersive GWs interacting with matter is outlined in \myRef{my:gwjeans} and, within a broader context, in \myRef{my:ql}. The work \cite{my:gwponder} also specifies some elementary blocks of our general theory, such as the so-called ponderomotive potential and the gravitational susceptibility. We discussed applications of our theory to GWs in neutral gases in \myRef{my:gwjeans} and non-magnetized plasmas in \myRef{tex:mygwplasma}, focusing mostly on linear dispersion relations. A QL theory of GWs has been proposed in \myRef{my:ql}, but it is gauge-dependent by construction and thus free of coordinate artifacts only asymptotically. We have also discussed the manifestly gauge-invariant approach to GWs in a general metric, but only in the context of a linear theory so far \cite{my:gwinvar}. Here, we merge those two approaches and, for the first time, formulate a manifestly gauge-invariant theory of dispersive GWs beyond the linear approximation.

Specifically, we show how to rewrite the linear-GW action, approximately yet without loss of accuracy, in terms of the gauge-invariant part (projection) of the metric perturbation that was derived in \myRef{my:gwinvar}. We accomplish this for a general background metric, so our approach is applicable to waves propagating in matter arbitrarily (but smoothly) distributed in spacetime. For this, we assume the short-wavelength limit, which allows us to bypass the problems \cite{ref:isi18, ref:caprini18, ref:riles13, ref:su12, ref:zalaletdinov96, tex:zalaletdinov97, ref:stein11, ref:green11, tex:green15, ref:buchert15, ref:kaspar12, ref:clarkson11, phd:kaspar14} associated with covariant self-consistent averaging on curved manifolds.\footnote{An alternative approach is presented in \myRef{my:ql} (which is not concerned with exact gauge invariance but includes nonadiabatic interactions). There, averaging is done in phase space, so the problems of averaging over a curved background \cite{ref:green11, tex:green15, ref:buchert15, ref:kaspar12, ref:clarkson11, phd:kaspar14} do not emerge, because phase space is a symplectic rather than a Riemannian manifold and has no concept of a metric to begin with.} Then, geometrical-optics (GO) equations are derived for linear GWs from the approximated action as usual \cite{book:tracy, my:quasiop1, my:amc}. This approach is commonly known to provide a particularly simple and robust way of deriving reduced equations \cite{ref:dougherty70, phd:ruiz17, my:itervar}, especially compared to averaging differential equations directly \cite{my:mquanta, my:dense}. Next, we identify the action and the energy--momentum of a dispersive GW in a gauge-invariant form and derive the nonlinear effect of GW on the background medium. We also show how gauge invariance can be maintained within a given accuracy if nonlinearities are included up to an arbitrary order in the GW amplitude.  An application to vacuum GWs is presented as a simple example that connects our general formulation with commonly known results. More importantly, though, our results show, for the first time, that a reduced QL theory of GWs in matter can be entirely freed of gauge artifacts and, as such, represents a physically meaningful framework to study specific GW modes in matter in the future.

Our paper is organized as follows. In \Sec{sec:prelim}, we introduce our basic equations. In \Sec{sec:variational}, we propose a reduced variational formulation for weak short-wavelength GWs on a smooth background spacetime and put it in a gauge-invariant form. In \Sec{sec:linear}, we introduce the equations for the perturbation metric and our gauge-invariant formulation of GO of GWs. We also discuss how vacuum GWs are described within this framework, and in particular, how their action and energy--momentum are naturally introduced in a gauge-invariant form. In \Sec{sec:QL}, we present the model for how weak dispersive GWs influence the background metric. We also discuss vacuum waves as an example. In \Sec{sec:euler}, we present our second approach to gauge-invariant GW theory that extends beyond the QL approximation. In \Sec{sec:sum}, we summarize our main results.

\section{Preliminaries}
\label{sec:prelim}

\subsection{Einstein equations}

Let us consider a metric $\total{g}_{\alpha\beta}$ with signature $(-+++)$ on a four-dimensional spacetime with coordinates  denoted as $x^\alpha$. The dynamics of this metric is governed by the least-action principle \cite{book:landau2}
\begin{gather}\label{eq:S}
\delta\tS = 0, 
\quad
\tS = \tS_{\rm m} + \tS_{\rm EH}.
\end{gather}
Here, $\tS_{\rm m}$ is the action of the matter, $\tS_{\rm EH}$ is the vacuum action of the gravitational field,
\begin{gather}\label{eq:SEH}
\tS_{\rm EH} = \frac{1}{2\kappa} \int \total{R}\,\sqrt{-\total{g}}\,\dd^4x,
\end{gather}
called the Einstein--Hilbert action, $\total{R}$ is the Ricci scalar, $\total{g} \doteq \det \total{g}_{\alpha\beta}$, and $\doteq$ denotes definitions. By default, we assume units such that the Einstein constant $\kappa \equiv 8\pi G/c^4$ and the speed of light $c$ are equal to unity,
\begin{gather}
c = 8\pi G = 1.
\end{gather}

The equations for $\total{g}_{\alpha\beta}$, called the Einstein equations, are obtained from
\begin{gather}\label{eq:Stot}
\sigma_{\alpha\beta}[\total{g}] \doteq \frac{\delta\tS[\total{g}]}{\delta \total{g}^{\alpha\beta}} = 0,
\end{gather}
where $\smash{\total{g}^{\alpha\beta}}$ is the inverse metric (\(\smash{\total{g}^{\alpha\beta}\total{g}_{\beta\gamma}} = \smash{\delta^\alpha_\gamma}\)) and $[\total{g}]$ denotes that the result is evaluated on $\smash{\total{g}^{\alpha\beta}}$. Using
\begin{gather}
\frac{\delta\tS_{\rm EH}[\total{g}]}{\delta \total{g}^{\alpha\beta}} 
= \frac{\sqrt{-\total{g}}}{2}\,\total{G}_{\alpha \beta},
\quad
\frac{\delta\tS_{\rm m}[\total{g}]}{\delta \total{g}^{\alpha\beta}} 
= -\frac{\sqrt{-\total{g}}}{2}\,\total{T}_{\alpha \beta},
\label{eq:efe}
\end{gather}
where $\total{G}_{\alpha \beta}$ is the Einstein tensor and $\total{T}_{\alpha \beta}$ is the energy-momentum tensor of the matter, \Eq{eq:Stot} can be represented~as
\begin{gather}
\label{eq:GT1}
\total{G}_{\alpha\beta} = \total{T}_{\alpha\beta}.
\end{gather}
We will assume, for clarity, that the matter is \textit{not} ultra-relativistic; then each element of $\total{T}_{\alpha \beta}$ is \(\mc{O}(\rho)\), where $\rho$ is the mass density.

\subsection{GW and average metric}
\label{sec:metric}

Here, we consider a class of problems within which the GW propagation is typically studied. Specifically, let us assume the existence of, and restrict our consideration to, the coordinates such that $\total{g}_{\alpha\beta}$ can be decomposed (in a way to be specified shortly) as
\begin{gather}
\total{g}_{\alpha\beta} = g_{\alpha\beta} + h_{\alpha\beta},
\label{eq:gtotal}
\end{gather}
where each element of $g_{\alpha\beta}$ is \(\mc{O}(a^0)\) and each element of $h_{\alpha\beta}$ has a magnitude that does not exceed a small constant $\hm \ll 1$. Furthermore, we assume that $h_{\alpha\beta}$ has two separate scales: $\ell_h$, which is the characteristic wavelength of GWs of interest, and $\ell_g \gg \ell_h$, which is the scale of the background or its radius of curvature.\footnote{Strictly speaking, the longer length scale does not need to be tied to the background inhomogeneity. It could also be the scale of the envelope, if the radius of the background curvature is larger. For simplicity, we do not elaborate on these details but rather focus on the class of problems that are typically considered in the context of GW propagation \cite{ref:isaacson68a, ref:madore73, ref:chesters73, ref:andersson21}.} Then, our definition of the scale separation is the same as that of the high-frequency limit in \myRef{ref:isaacson68a}, where a detailed discussion about its physical relevance can also be found.

Because of the assumed scale separation, one can find a scale $\ell_a$ that satisfies
\begin{subequations}\label{eq:golim}
\begin{gather}
\ell_h \ll \ell_a \ll \ell_g,
\label{eq:scalesep}
\end{gather}
and one can introduce a small ``GO parameter''\footnote{Note that the applicability of the GO approximation to dispersive GWs is generally more restricted than that for other waves, for example, electromagnetic waves in dispersive media \cite{my:gwjeans}.}
\begin{gather}
\epsilon \doteq \ell_h/\ell_g = (\ell_h/\ell_a)^2 = (\ell_a/\ell_g)^2 \ll 1.
\end{gather}
\end{subequations}
More formally, we assume that any $h_{\alpha\beta}$ of interest is a superposition of quasiperiodic functions, \ie functions of $(\epsilon x, \theta(x))$, where $\epsilon \equiv (\ell_h/\ell_a)^2$ is a small dimensionless parameter and the dependence on $\theta$ is $2\pi$-periodic. Then, any function $A$ (generally, a tensor) that contains $\total{g}_{\alpha\beta}$ is also quasiperiodic with scales that satisfy \eq{eq:golim} and can be assigned a local average $\favr{\ldots}$ over a spacetime volume of size $\ell_a$ as
\begin{gather}
\favr{A}(x) \doteq \int \dd^4 x'\,\Psi(x, x') A(x').
\end{gather}
Here, $\Psi$ is a fixed function that is localized on the scale $\ell_a$ near $x' = x$ and is normalized such that $\int \dd^4 x'\,\Psi(x, x') = 1$. The shape of $\Psi$ can be arbitrary, because of the following. As a quasiperiodic function, $A$ can be represented as a Fourier series in~$\theta$:
\begin{gather}
A = \sum_n A_n(\epsilon x) \ee^{\ii n \theta(x)}.
\label{eq:AA}
\end{gather}
Then, $\smash{\favr{A_n(\epsilon x) \ee^{\ii n \theta(x)}}}$ is exponentially small with respect to $\sqrt{\epsilon}$ except for $n = 0$, and $\smash{\favr{A_0(\epsilon x)}} \approx A_0(\epsilon x)$ with error of order $\sqrt{\epsilon}$, because on the scale $\ell_g$, the function $\Psi$ is close to delta function. This means that $\favr{A} \approx A_0$, which becomes an exact equality at $\epsilon \to 0$ independently of the shape of $\Psi$ and $\ell_a$ as long as \Eq{eq:golim} is satisfied. (Corrections caused by nonzero $\epsilon$ will not be important for our purposes.)  This subject is discussed in detail in \myRef{my:ql}. Furthermore, notice the following. The aforementioned restriction on the choice of background coordinates implies that any coordinate transformation that is \(\mc{O}(a^0)\) is assumed to have the scale \(\ell_g\) or larger. [See also \Eqs{eq:dhtr} and the ensuing discussion.] This means that, at $\epsilon \to 0$, such transformations preserve the expansion \eq{eq:AA} in that each $A_n$ remains the $n$th Fourier harmonic of $A$ with respect to the phase $\theta$, which is a true scalar. Then, each coefficient $A_n$ transforms as a tensor independently from the other coefficients in the expansion \eq{eq:AA}, and thus, in particular, our averaging procedure is frame-invariant at $\epsilon \to 0$.  (It is also possible to use other averaging schemes \cite{ref:brill64, ref:zalaletdinov96, tex:zalaletdinov97}, and those are known to produce equivalent results \cite{ref:isi18, ref:caprini18, ref:riles13, ref:su12,ref:stein11} under the condition \eq{eq:scalesep}.)

We now specify the splitting \eq{eq:gtotal} by requiring that $h_{\alpha\beta}$ satisfies the following condition that is invariant with respect to allowed background transformations at $\epsilon \to 0$:
\begin{gather}\label{eq:havr}
\favr{h_{\alpha\beta}} = 0.
\end{gather}
We call such a perturbation a GW. Then, \(g_{\alpha\beta}\) can be understood as the background metric for the GW or as the average part of the total metric:
\begin{gather}\label{eq:gavr}
g_{\alpha\beta} = \favr{\total{g}_{\alpha\beta}}.
\end{gather}
For any pair of (complex) fields $u_1$ and $u_2$ on the background space, we introduce the following inner product:
\begin{gather}\label{eq:inner}
\braket{u_1, u_2} = \int \dd^4 x\,\sqrt{-g}\,u_1^*(x) u_2(x),
\end{gather}
where $g \doteq \det g_{\alpha\beta}$. We also introduce the inverse background metric $g^{\alpha\beta}$ via $g^{\alpha\beta}g_{\beta\gamma} = \delta^\alpha_\gamma$. Then the inverse total metric  \(\total{g}^{\alpha\beta}\) can be expressed as follows:
\begin{gather}
\total{g}^{\alpha\beta} = g^{\alpha\beta} - h^{\alpha\beta} + {h^\alpha}_\gamma h^{\gamma\beta} + \mc{O}(\hm^3).
\label{eq:hupper}
\end{gather}
(Here and further, the indices of the perturbation metric $h_{\alpha\beta}$ are manipulated using the background metric and its inverse, unless specified otherwise.) We will assume that the matter responds adiabatically to GWs, so $g_{\alpha\beta}$ and $h_{\alpha\beta}$ are the only degrees of freedom, meaning that resonant interactions of waves with matter are ignored (but see \myRef{my:ql} for these effects). This is a standard approach that was used in the past for electromagnetic and other interactions, for example, in \Refs{my:qponder, ref:dewar77, ref:cary81, my:gwponder, my:acti, my:sharm} and is reviewed, for example, in \myRef{my:itervar}.

The sign conventions will be assumed as in \Refs{book:carroll, book:misner77}. Then, in particular, the commutator of two covariant derivatives acting on any vector field $u^\alpha$ can be expressed~as
\begin{gather}\label{eq:Ru}
[\del_\beta, \del^\alpha]u^\beta = R^\alpha{}_\beta u^\beta,
\end{gather}
where $R_{\alpha\beta} \equiv \smash{g_{\alpha\gamma}R^\gamma{}_\beta}$ is the the Ricci tensor of the background metric. In addition to these, we also introduce the operator 
\begin{gather}
\oper{Q}^\alpha{}_\beta \doteq - \delta_\beta^\alpha \del_\mu \del^\mu - R^\alpha{}_\beta,
\label{eq:QX}
\end{gather}
and we define $\smash{\oper{\Xi}^\alpha{}_\beta}$ as the Green's operator of the following equation for $\smash{u^\alpha}$:
\begin{gather}\label{eq:Q}
\oper{Q}^\alpha{}_\beta u^\beta = q^{\alpha},
\end{gather}
where $q^{\alpha}$ is a given vector field. In other words, $\smash{\oper{\Xi}^\alpha{}_\beta}$ is such that
\begin{gather}\label{eq:xisol}
u^\alpha = \oper{\Xi}^\alpha{}_\beta q^\beta.
\end{gather}

Notably, \(\oper{Q}^\alpha{}_\beta\) is a hyperbolic operator similar to the one from the driven Maxwell's equation for the Lorenz-gauge in vacuum \cite{my:spinhall} except for the opposite sign in front of the Ricci tensor. We will assume the adiabatic limit, when $\smash{\oper{Q}^\alpha{}_\beta}$ is approximately invertible due to the scale separation \eq{eq:golim}, in which case one can write
\begin{gather}
\oper{\Xi}^\alpha{}_\beta \approx (\oper{Q}^{-1})^\alpha{}_\beta.
\label{eq:xiadiab}
\end{gather}
Further details can be found in \myRef{my:gwinvar}.

\section{Variational approach}
\label{sec:variational}

A general approach to QL theory involves the Weyl calculus, as detailed in \myRef{my:ql}. However, it requires cumbersome machinery that is redundant for describing adiabatic waves that we are interested in here. For the purposes of this paper, it is sufficient to use a simpler and more intuitive Whitham's method, as motivated in \Sec{sec:intro}. This is the approach we adopt below. That said, readers are also encouraged to consider \myRef{my:ql} as an affirmation of Whitham's method for adiabatic GWs (in particular, see Sec.~9.4 there), even though \myRef{my:ql} is not concerned with exact gauge invariance that we pursue here. 

\subsection{Basic equations}

Using \Eq{eq:hupper} for the inverse total metric, the total action $\tS$ can be expanded in $h^{\alpha\beta}$ as follows:
\begin{subequations}\label{eq:Sder}
\begin{gather}
\tS[\total{g}] = \tS[g] + S^{(1)}[g, h]+ S^{(2)}[g, h] + \mc{O}(\hm^3),\\
S^{(1)}[g, h] = - \braket{\sigma_{\alpha\beta}[g], h^{\alpha\beta}},\\
S^{(2)}[g, h] = \frac{1}{2}\braket{
h^{\alpha\beta}, \oper{\total{D}}_{\alpha\beta\gamma\delta}[g]h^{\gamma\delta}
}.\label{eq:SD}
\end{gather}
\end{subequations}
Here, the square brackets $[g]$ emphasize that the corresponding expansion coefficients, such as the matrix function $\smash{\sigma_{\alpha\beta}}$ and the operator $\smash{\oper{\total{D}}_{\alpha\beta\gamma\delta}}$, are evaluated on $g^{\alpha\beta}$. Also, without loss of generality, one can assume that $\smash{\sigma_{\alpha\beta}} = \smash{\sigma_{\beta\alpha}}$ and
\begin{subequations}\label{eq:Dsym}
\begin{gather}
(\oper{\total{D}}_{\alpha\beta\gamma\delta})^\dag
=\oper{\total{D}}_{\gamma\delta\alpha\beta},
\label{eq:Dsym1}
\\
\oper{\total{D}}_{\alpha\beta\gamma\delta} 
= 
\oper{\total{D}}_{\beta\alpha\gamma\delta} 
=
\oper{\total{D}}_{\alpha\beta\delta\gamma}
=
\oper{\total{D}}_{\beta\alpha\delta\gamma},
\end{gather}
\end{subequations}
where the dagger denotes Hermitian adjoint with respect to the inner product \eq{eq:inner}.

Because $\smash{\favr{h^{\alpha\beta}}} = 0$, the linear term in \Eq{eq:Sder} vanishes. [To reiterate, this is possible only within the short-wavelength approximation \eq{eq:golim}.] Let us also assume for clarity that the three-wave interactions are negligible (a sufficient condition for this is that the GW spectrum must be not too broad); then the average of terms cubic in $\smash{h^{\alpha\beta}}$ vanishes too \cite[Sec.\ 1.1.2]{book:zakharov-b}. This leads~to
\begin{gather}\label{eq:S22}
\tS[\total{g}] = \tS[g] + S^{(2)}[g, h] + \Delta S[g, h],
\end{gather}
where $\Delta S = \smash{\mc{O}(\hm^4)}$. The term $\smash{S^{(2)}}$ can be represented as a sum of the vacuum action $\smash{S^{(2)}_{\rm vac}}$ and the term $\smash{S^{(2)}_{\rm m}} $ that describes the GW--matter coupling:
\begin{gather}
\smash{S^{(2)}} = 
\underbrace{{S^{(2)}_{\rm vac}}}_{\mc{O}(\hm^2)} 
+ 
\underbrace{{S^{(2)}_{\rm m}}}_{\mc{O}(\rho\hm^2)}.
\end{gather}
Assuming $\smash{\hm^2} = o(\rho)$, the term $\Delta S$ in \Eq{eq:S22} is small not only with respect to $\smash{S^{(2)}_{\rm vac}}$, but also with respect to $\smash{S^{(2)}_{\rm m}}$. (The formulation presented below is valid also at arbitrarily small $\rho$ to the extent that the GW interactions with matter can be neglected.) Then, the system can be described with the following truncated action that retains not only the vacuum linearized gravity but the GW--matter interactions too:
\begin{gather}\label{eq:truncS}
\tS[\total{g}] = \tS[g] + S^{(2)}[g, h].
\end{gather}
An example of $\smash{S^{(2)}[g, h]}$ that emerges from the quasimonochromatic-GW interaction with neutral gas was derived in \myRef{my:gwponder} as a part of studying the ponderomotive effect of GWs on matter.

\subsection{Coordinate transformations}
\label{sec:ct}

Let us explore how the action \eq{eq:S22} is transformed under near-identity coordinate transformations
\begin{gather}\label{eq:vt}
x^\mu \to x'^\mu \doteq x^\mu + \xi^\mu, \quad \xi^\mu = \mc{O}(a),
\end{gather}
where $\xi^\mu$ is a small vector field. A transformation \eq{eq:vt} induces the following transformation of the total metric:
\begin{gather}
\total{g}^{\mu\nu} \to \total{g}'^{\mu\nu} =
\total{g}^{\mu\nu} - \lie_\xi \total{g}^{\mu\nu} + \mc{O}(\hm^2),
\end{gather}
where $\smash{\lie_\xi}$ is the Lie derivative along~$\smash{\xi^\mu}$~\cite{book:carroll},
\begin{gather}\label{eq:lie}
\lie_\xi g^{\alpha\beta} = -\del^\alpha \xi^\beta - \del^\beta \xi^\alpha.
\end{gather}

Let us specifically focus on the case $\favr{\xi^\mu} = 0$. Then, using \Eq{eq:hupper} together with \Eqs{eq:havr} and \eq{eq:gavr}, one obtains the following transformation of the perturbation metric, or a \textit{gauge transformation} with $\xi^\mu$ serving as a gauge field:
\begin{gather}\label{eq:htr}
h^{\mu\nu} \to h'^{\mu\nu} =
h^{\mu\nu} + \lie_\xi g^{\mu\nu} + \mc{O}(\hm^2).
\end{gather}
Also, the background metric is transformed accordingly as $\smash{g_{\mu\nu}} \to \smash{g'_{\mu\nu}} = \favr{\total{g}'_{\mu\nu}}$, so
\begin{gather}\label{eq:Delg}
\Delta g^{\mu\nu} \doteq g'^{\mu\nu} - g^{\mu\nu} = \mc{O}(\hm^2).
\end{gather}

From \Eq{eq:truncS}, $\tS[\total{g}']$ can be written as
\begin{gather}
\tS[\total{g}'] =
\tS[g']
+ \frac{1}{2}\braket{
h'^{\alpha\beta}, \oper{\total{D}}_{\alpha\beta\gamma\delta}[g]h'^{\gamma\delta}
}
+ \mc{O}(\hm^4),
\label{eq:SS0}
\end{gather}
where the difference between $\smash{\oper{\total{D}}_{\alpha\beta\gamma\delta}[g']}$ and $\smash{\oper{\total{D}}_{\alpha\beta\gamma\delta}[g]}$ has been absorbed into $\smash{\mc{O}(\hm^4)}$. Because $\tS[\total{g}'] = \tS[\total{g}]$ by general covariance, this leads~to
\begin{multline}\label{eq:SS}
\tS[g'] - \tS[g]
+ \braket{
h^{\alpha\beta}, \oper{\total{D}}_{\alpha\beta\gamma\delta}[g]\lie_\xi g^{\gamma\delta}
}
\\ + \frac{1}{2}\braket{
\lie_\xi g^{\alpha\beta}, \oper{\total{D}}_{\alpha\beta\gamma\delta}[g]\lie_\xi g^{\gamma\delta}
}
= \mc{O}(\hm^4),
\end{multline}
where we used \Eqs{eq:Dsym1} and \eq{eq:htr}. Notice that
\begin{gather}
\tS[g'] - \tS[g]
\approx \braket{\sigma_{\alpha\beta}[g], \Delta g^{\mu\nu}} = \mc{O}(\hm^4),
\end{gather}
where we used \Eq{eq:Delg} and the fact that $\sigma_{\alpha\beta}[g] = \mc{O}(\hm^2)$ by the Einstein equations. After substituting this into \Eq{eq:SS}, one obtains
\begin{multline}\label{eq:aux1}
\braket{h^{\alpha\beta}, \oper{\total{D}}_{\alpha\beta\gamma\delta}[g]\lie_\xi g^{\gamma\delta}}
\\+ \frac{1}{2}\braket{
\lie_\xi g^{\alpha\beta}, \oper{\total{D}}_{\alpha\beta\gamma\delta}[g]\lie_\xi g^{\gamma\delta}
}
= \mc{O}(\hm^4).
\end{multline}
Because $\xi^\mu$ is arbitrary, one can also flip the sign of $\xi^\mu$ to obtain
\begin{multline}\label{eq:aux2}
-\braket{h^{\alpha\beta}, \oper{\total{D}}_{\alpha\beta\gamma\delta}[g]\lie_\xi g^{\gamma\delta}}
\\+ \frac{1}{2}\braket{
\lie_\xi g^{\alpha\beta}, \oper{\total{D}}_{\alpha\beta\gamma\delta}[g]\lie_\xi g^{\gamma\delta}
}
= \mc{O}(\hm^4).
\end{multline}
Together, \Eqs{eq:aux1} and \eq{eq:aux2} lead to
\begin{subequations}\label{eq:Dhx}
\begin{gather}
\braket{h^{\alpha\beta}, \oper{\total{D}}_{\alpha\beta\gamma\delta}[g]\lie_\xi g^{\gamma\delta}} = \mc{O}(\hm^4),\\
\braket{
\lie_\xi g^{\alpha\beta}, \oper{\total{D}}_{\alpha\beta\gamma\delta}[g]\lie_\xi g^{\gamma\delta}
} = \mc{O}(\hm^4).
\end{gather}
\end{subequations}
As a side remark, note that within the linear approximations, the \(\mc{O}(a^4)\) terms in \Eqs{eq:SS0} [and the \(\mc{O}(a^2)\) term in \Eq{eq:Delg}] are negligible. Then the right-hand side of \Eqs{eq:Dhx} is zero, which shows that linear theory is covariant under transformations $\smash{h^{\mu\nu} \to h'^{\mu\nu} = h^{\mu\nu} + \lie_\xi g^{\mu\nu}}$, or \textit{gauge-invariant}.

In contrast, the reduced field theory induced by the approximate action \eq{eq:truncS} is not strictly gauge-invariant; \ie $\tS$ generally \textit{does} change under coordinate transformations \eq{eq:vt}. These changes are $\smash{\mc{O}(\hm^4)}$, so they are beyond the accuracy of our approximation and therefore do not invalidate the reduced theory \textit{per~se}. Still, the lack of covariance makes the reduced theory not entirely satisfactory, as discussed in \Sec{sec:intro}. Below, we show how to derive an alternative approximation of $\tS$ that, on one hand, is equivalent to \Eq{eq:truncS} within the accuracy of our theory but, on the other hand, is exactly gauge-invariant.

\subsection{Projecting on the invariant subspace}

As shown in our recent \myRef{my:gwinvar}, the perturbation metric can be uniquely decomposed into the gauge-invariant part $\smash{\psi^{\alpha\beta}}$ and the remaining gauge part that is the Lie derivative of some vector field $\zeta^\alpha$:
\begin{subequations}\label{eq:decomp}
\begin{gather}
h^{\alpha\beta} = \psi^{\alpha\beta} + \lie_\zeta g^{\alpha\beta},\\
\psi^{\alpha\beta} = \oper{\Pi}^{\alpha\beta} {}_{\gamma\delta} h^{\gamma\delta}.
\label{eq:psi}
\end{gather}
\end{subequations}
(The notation here is slightly different than in \myRef{my:gwinvar}.) Specifically, 
\begin{gather}
\zeta^\alpha \doteq \oper{\Xi}^{\alpha}{}_{(\gamma} \del_{\delta)}h^{\gamma\delta}
- \frac{1}{2}\,\oper{\Xi}^{\alpha}{}_{\mu} \del^\mu g_{\gamma\delta}h^{\gamma\delta},
\label{eq:zeta}
\end{gather}
where the linear operator $\smash{\oper{\Xi}^\alpha{}_\beta}$ is defined in \Eq{eq:xiadiab}. Also, the linear operator $\smash{\oper{\Pi}^{\alpha\beta}{}_{\gamma\delta}}$ is given by
\begin{gather}
\oper{\Pi}^{\alpha\beta}{}_{\gamma\delta} 
=
\delta^\alpha_{(\gamma}\delta^\beta_{\delta)}
+2\del^{(\alpha} \oper{\Xi}^{\beta)}{}_{(\gamma} \del_{\delta)}
-\del^{(\alpha} \oper{\Xi}^{\beta)}{}_{\mu} \del^\mu g_{\gamma\delta},
\end{gather}
and has the following properties:
\begin{subequations}\label{eq:props}
\begin{gather}
\oper{\Pi}^{\alpha\beta} {}_{\gamma\delta}
\oper{\Pi}^{\gamma\delta} {}_{\lambda\varepsilon} 
= \oper{\Pi}^{\alpha\beta} {}_{\lambda\varepsilon},
\label{eq:proj1}
\\
\oper{\Pi}^{\alpha\beta} {}_{\gamma\delta} \lie_u g^{\gamma\delta} = 0,
\label{eq:PiLie}
\\
\oper{\Pi}^{\alpha\beta} {}_{\gamma\delta}
= \oper{\Pi}^{\alpha\beta} {}_{\delta\gamma}
= \oper{\Pi}^{\beta\alpha} {}_{\gamma\delta},
\label{eq:Pisym}
\end{gather}
\end{subequations}
where $u^\mu$ is any vector field. 

Substituting \Eqs{eq:decomp} into \Eq{eq:SD} for $\smash{S^{(2)}}$ yields
\begin{align}
S^{(2)} = & \,
\frac{1}{2}\braket{\psi^{\alpha\beta}, \oper{\total{D}}_{\alpha\beta\gamma\delta}[g]\psi^{\gamma\delta}}
\notag\\
& + \braket{\psi^{\alpha\beta}, \oper{\total{D}}_{\alpha\beta\gamma\delta}[g]\lie_\zeta g^{\gamma\delta}}
\notag\\
& + \frac{1}{2}\braket{\lie_\zeta g^{\alpha\beta}, \oper{\total{D}}_{\alpha\beta\gamma\delta}[g]\lie_\zeta g^{\gamma\delta}},
\end{align}
where we used \Eqs{eq:Dsym}. By \Eqs{eq:Dhx}, the second and the third term here are $\mc{O}(\hm^4)$ and therefore negligible within the accuracy of our theory. Hence, only the first term should be retained. Using \Eq{eq:psi}, one thereby arrives at the following approximation: 
\begin{subequations}\label{eq:Sappr}
\begin{gather}
\tS = \tS[g] + S^{(2)}[g, h],\label{eq:tS}\\
S^{(2)} =
\frac{1}{2}\braket{\oper{\Pi}^{\alpha\beta} {}_{\mu\nu} h^{\mu\nu}, 
\oper{\total{D}}_{\alpha\beta\gamma\delta}[g]\,
\oper{\Pi}^{\gamma\delta} {}_{\varepsilon\lambda} h^{\varepsilon\lambda}},
\end{gather}
\end{subequations}
where we have repeated \Eq{eq:truncS} for completeness. Equation \eq{eq:Sappr} defines a field theory where $g_{\alpha\beta}$ and $h_{\alpha\beta}$ are independent fields. The advantage of this approximate field theory is that it is \textit{exactly} invariant under gauge transformations
\begin{gather}\label{eq:hpp0}
h^{\alpha\beta} \to h'^{\alpha\beta} \doteq h^{\alpha\beta} + \lie_\xi g^{\alpha\beta}
\end{gather}
for any $\xi^\mu$. Specifically, by \Eq{eq:PiLie}, one has
\begin{gather}
\tS[g, h'] = \tS[g, h].
\end{gather}

Let us also consider a class of transformations $g_{\alpha\beta} \to g'_{\alpha\beta}$ and $h_{\alpha\beta}\to h'_{\alpha\beta}$ such that
\begin{subequations}\label{eq:dhtr}
\begin{gather}
g'_{\alpha'\beta'} \doteq \frac{\pd x^\alpha}{\pd x'^{\alpha'}}\frac{\pd x^\beta}{\pd x'^{\beta'}}\,g_{\alpha\beta},
\\
h'_{\alpha'\beta'} \doteq \frac{\pd x^\alpha}{\pd x'^{\alpha'}}\frac{\pd x^\beta}{\pd x'^{\beta'}}\,h_{\alpha\beta},
\end{gather}
\end{subequations}
where \(x = x(x')\) is a prescribed four-dimensional function. This can be understood as the zeroth-order coordinate transformation on the original spacetime (or a regular transformation of the background coordinates). Since  \(\oper{\Pi}^{\alpha\beta} {}_{\gamma\delta}[g]\) is defined covariantly with respect to such transformations, the object \(\oper{\Pi}^{\gamma\delta} {}_{\varepsilon\lambda} h^{\varepsilon\lambda}\) is a true tensor on it. Similarly, covariance of \(\oper{\total{D}}_{\alpha\beta\gamma\delta}[g]\) and $S[g]$ ensures that \(S^{(2)}\) and the whole action \eq{eq:tS} are true scalars on the background spacetime. Thus, in addition to the gauge symmetry, the field theory \eq{eq:Sappr} is also invariant with respect to transformations \eq{eq:dhtr}.

\subsection{Metric perturbation as a vector field}
\label{sec:latin}

Following \myRef{my:gwinvar}, let us consider $h^{\alpha\beta}$ as a 16-dimensional field $h^a$, or $\vec{h}$ in the index-free notation, of the form
\begin{gather}
\vec{h} = (h^{00}, h^{01}, h^{02}, h^{03}, h^{10}, \ldots, h^{32}, h^{33})^\intercal,
\label{eq:perturbvec}
\end{gather}
where $^\intercal$ denotes transpose. (Here and further, Latin indices from the beginning of the alphabet range from 1 to 16.) In other words,
\begin{align}
h^a = h^{\alpha\beta}, & \quad h_b = h_{\gamma\delta},\\
\{\alpha,\beta\} = \iota(a), & \quad \{\gamma,\delta\} = \iota(b),\label{eq:iot0}
\end{align}
where the index function $\iota$ is defined via
\begin{gather}
\iota(a) \doteq \big\{
1 + \lfloor (a-1)/4 \rfloor,
1 + (a-1)\,\text{mod}\,4
\big\},
\end{gather}
assuming $a,b = 1, 2, \ldots, 16$. Accordingly,
\begin{gather}\label{eq:Gamma}
h_b = \gamma_{ba} h^a, 
\quad
\gamma_{ab} \doteq g_{\gamma\alpha}g_{\delta\beta},
\end{gather}
assuming the same notation as in \Eq{eq:iot0}.

Let us define $\mcu{H}_1$ as a Hilbert space of one-component functions on the background spacetime with the usual inner product $\braket{\cdot\,, \cdot}$. Then, the 16-dimensional fields~\eq{eq:perturbvec} can be considered as vectors in the Hilbert space $\mcu{H}_{16}$ that is the tensor product of 16 copies of $\mcu{H}_1$, with the inner product
\begin{gather}\label{eq:inner2}
\braket{\vec{\xi}| \vec{\varphi}} = \sum_{a=1}^{16} \braket{\xi_a, \varphi^a},
\end{gather}
where $\braket{\cdot\,, \cdot}$ is given by \Eq{eq:inner}. (Unlike in the rest of the paper, summation is shown explicitly here in order to emphasize the difference between $\braket{\cdot\,|\, \cdot}$ and $\braket{\cdot\,, \cdot}$.) Then, $\smash{\oper{\Pi}^{\alpha\beta}{}_{\gamma\delta}}$ induces an operator $\smash{\oper{\Pi}^a{}_b}$ on $\mcu{H}_{16}$ defined via
\begin{gather}\label{eq:Pidef}
(\boper{\Pi}\vec{h})^a \equiv \oper{\Pi}^a{}_b h^b \doteq \oper{\Pi}^{\alpha\beta}{}_{\gamma\delta} h^{\gamma\delta}.
\end{gather}
By \Eq{eq:proj1}, one has
\begin{gather}\label{eq:Pipr}
\boper{\Pi}{}^2 = \boper{\Pi},
\quad
\boper{\Pi}\vec{\phi} = 0,
\end{gather}
where $\phi^{\alpha\beta} \doteq \lie_\xi g^{\alpha\beta}$ and $\xi^\mu$ is any vector field. This shows that $\smash{\boper{\Pi}}$ is a projector of the metric perturbation on the gauge-invariant subspace. However, note that $\smash{\boper{\Pi}{}^\dag \ne \boper{\Pi}}$, so $\smash{\boper{\Pi}}$ is not an orthogonal but oblique projector. [Here, the dagger denotes Hermitian adjoint with respect to the inner product \eq{eq:inner2}.] Using this machinery, one can rewrite \Eqs{eq:Sappr} in the following compact form:
\begin{subequations}\label{eq:Sapprv}
\begin{align}
\tS[\total{g}] 
& = \tS[g] + \tS^{(2)}[g, \vec{h}],
\\
\tS^{(2)}[g, \vec{h}]
& = \frac{1}{2}\braket{\vec{h} | \boper{D}[g] \vec{h}} \\
& \equiv \frac{1}{2}\braket{h^a , \oper{D}_{ab}[g] h^{b}} \\
& \equiv \frac{1}{2}\braket{h^{\alpha\beta}, \oper{D}_{\alpha\beta\gamma\delta}[g] h^{\gamma\delta}}.
\label{eq:S2short}
\end{align}
Here, we have introduced the operator
\begin{gather}\label{eq:Ddef}
\boper{D} \doteq \boper{\Pi}{}^\dag\boper{\total{D}} \boper{\Pi},\\
\oper{D}_{ab} = (\oper{\Pi}^c{}_a)^\dag \oper{\total{D}}_{cd} \oper{\Pi}^d{}_b 
= (\oper{\Pi}^\dag)_a{}^c \oper{\total{D}}_{cd} \oper{\Pi}^d{}_b,
\end{gather}
\end{subequations}
which is Hermitian by \Eqs{eq:Dsym}. The indices here can be raised and lowered using $\gamma_{ab}$ [\Eq{eq:Gamma}] as a metric.

\subsection{Metric perturbation as a complex variable}

For dynamics governed by \Eq{eq:truncS} to be consistent with the assumption of slow $\smash{g^{\alpha\beta}}$, the integrand in $\tS^{(2)}$ must be properly averaged. This is done as follows. Because $\smash{h^{\alpha\beta}}$ is assumed to be rapidly oscillating, one can unambiguously split it into the part $\smash{\tilde{h}^{\alpha\beta}}$ that corresponds to positive frequencies and the complex-conjugate part $\smash{\tilde{h}^{\alpha\beta*}}$ that corresponds to negative frequencies:
\begin{gather}\label{eq:comph}
h^{\alpha\beta} = \frac{1}{2}\,(\tilde{h}^{\alpha\beta} + \tilde{h}^{\alpha\beta*}),
\end{gather}
which is always doable provided the scale separation \eq{eq:golim} \cite{ref:brizard93}. Also, one can write $\smash{\tilde{h}^{\alpha\beta}}$ as a sum (possibly, integral) over all quasimonochromatic waves present in the system:
\begin{gather}
\tilde{h}^{\alpha\beta} = \sum_s \tilde{h}_s^{\alpha\beta},
\quad
\tilde{h}_s^{\alpha\beta} = a_s^{\alpha\beta} \ee^{\ii\theta_s},
\end{gather}
where the envelopes $\smash{a_s^{\alpha\beta}}$ and the local wavevectors $\bar{k}_\mu \doteq \pd_\mu \theta_s$ are slow compared with $\theta_s$ and the sum is taken over all the waves that are present in the system. In \Eq{eq:S2short}, contributions of $\smash{\tilde{h}^{\alpha\beta}\tilde{h}^{\gamma\delta}}$ and $\smash{\tilde{h}^{*\alpha\beta}\tilde{h}^{*\gamma\delta}}$ average to zero, and so are terms like $\smash{\tilde{h}_{s'}^{*\alpha\beta}\tilde{h}_s^{\gamma\delta}}$ with $s' \ne s$. Then,
\begin{gather}
\tS^{(2)}
= \frac{1}{4}\,\Re\sum_s\int \dd^4 x\sqrt{-g}\,\tilde{h}_s^{*\alpha\beta}\,\oper{D}_{\alpha\beta\gamma\delta} \tilde{h}_s^{\gamma\delta}.
\end{gather}
Using Hermiticity of $\smash{\boper{D}}$, one can drop ``$\Re$'' and rewrite $S_{\rm QL} \doteq S[\total{g}]$ as follows:
\begin{subequations}\label{eq:Savr0}
\begin{align}
S_{\rm QL} & = \tS[g] + S^{(2)}[g, \tilde{\vec{h}}_s, \tilde{\vec{h}}{}_s^*],
\\
S^{(2)}[g, \tilde{\vec{h}}_s, \tilde{\vec{h}}{}_s^*] & = \frac{1}{4} \sum_s \braket{\tilde{\vec{h}}_s| \boper{D}[g]\, \tilde{\vec{h}}_s}.
\label{eq:S2ql}
\end{align}
\end{subequations}

\subsection{Continuum wave spectrum}

As usual \cite{ref:kaufman87b, my:ql}, in the case of a continuum wave spectrum, $\smash{\tS^{(2)}}$ can be expressed through the spectrum of the correlation matrix of $\vec{h}$:
\begin{multline}\label{eq:wigner}
W^{ab}(x, k) \doteq \frac{1}{(2\pi)^4} \int \dd^4s\,\msf{J}(x, s)\,\exp(-\ii k_\nu s^\nu)
\\
\times \favr{h^a(x + s/2)h^b(x - s/2)},
\end{multline}
where \(s\) is a 4-tuple (not a vector), so \(x \pm s/2\) is to be understood as the index-free notation for \(x^\alpha \pm s^\alpha/2\), and $h^a(x \pm s/2)$ is the metric perturbation as a 16-dimensional vector field $h^a$ evaluated at the coordinates given by $x \pm s/2$. Note that \(W^{ab}\) is not a truly tensorial object in the general case. [See the discussion following \Eq{eq:s2wig1}.] It is also known as the (average) Wigner matrix and can be understood as the (average) Weyl symbol of the ``density operator'' \(\ket{\vec{h}}\bra{\vec{h}}\), at least up to a constant coefficient \cite{book:tracy, my:ql}. Here, $\msf{J}$ is a metric factor given by \cite{foot:supp, ref:gneiting13}
\begin{gather}\label{eq:J}
\msf{J}(x, s) \doteq [g(x + s/2)\,g(x - s/2)]^{1/4}.
\end{gather}
As a reminder, \(g(x)\) is the determinant of the background metric evaluated at~$x$. Much like in \Refs{ref:kaufman87b, my:ql}, one obtains
\begin{gather}
\tS^{(2)} = \frac{1}{2} \int \dd^4x\,\dd^4k \,D_{ab}(x, k)\, W^{ba}(x, k),
\end{gather}
where $D_{ab}$ is the Weyl symbol of $\oper{D}_{ab}$. In the index-free form, this can be written as
\begin{gather}\label{eq:s2wig1}
\tS^{(2)} = \frac{1}{2} \int \dd^4x\,\dd^4k \,\vec{D}(x, k):\vec{W}(x, k).
\end{gather}

As discussed in \myRef{my:quasiop1}, the matrices $D_{ab}$ and $W^{ab}$ are not necessarily true tensors. However, notice the following. In the GO limit, for any quasimonochromatic wave with a given local wavevector $k_\alpha$, one has \(\boper{D}\vec{h} \to \vec{D}(x, k)\vec{h}\). Since \(\boper{D}\) is a covariant object, \(\vec{D}\) is thereby a true tensor in this limit. Likewise, $W^{ab}$ becomes a tensor density in the GO limit. This is, in particular, because typical \(s^\alpha\) that contribute to the integral in \Eq{eq:wigner} are of the order of the correlation length (which in the GO limit is much less than \(\ell_g\)), so \(\msf{J}(x, s)\) can be replaced \(\sqrt{-g(x)}\). Then, $s$ can be approximately interpreted as a vector and $x \pm s/2$ can be understood as a small displacement of $x$ along this vector. More details about Wigner matrices and their role in QL theory can be found in \myRef{my:ql} and references cited therein. We are not focusing on this topic here because it is not needed for the rest of our paper, but see also \Sec{sec:qlvac}.

\section{Equations for the perturbation metric}
\label{sec:linear}

\subsection{Basic equations}

First, let us consider linear theory, where $\smash{g^{\alpha\beta}}$ is treated as a prescribed function. In this case, the field equations yielded by the action \eq{eq:Savr0} are as follows:\footnote{Instead of $\smash{(\tilde{h}_s^{\alpha\beta}, \tilde{h}_s^{\alpha\beta*})}$, one can as well use $\smash{(\tilde{h}_s^{\alpha\beta}, \tilde{h}_{s\alpha\beta}^*)}$ for the independent variables.}
\begin{gather}\label{eq:slieeq1}
\frac{\delta S}{\delta \tilde{h}_s^{\alpha\beta}} = 0,
\quad
\frac{\delta S}{\delta \tilde{h}_s^{\alpha\beta*}} = 0.
\end{gather}
These two equations are the complex conjugates of each other, so only one of them will be considered. Specifically, one obtains
\begin{subequations}\label{eq:Dh0}
\begin{gather}
\oper{D}_{\alpha\beta\gamma\delta} \tilde{h}_s^{\gamma\delta} = 0,
\end{gather}
so $\boper{D}$ is understood as the dispersion operator for GWs. Also note that because this equation is linear, one can equally write it as
\begin{gather}\label{eq:Dh1}
\oper{D}_{\alpha\beta\gamma\delta} \tilde{h}^{\gamma\delta} = 0
\end{gather}
and also as
\begin{gather}\label{eq:Dh}
\oper{D}_{\alpha\beta\gamma\delta} h^{\gamma\delta} = 0.
\end{gather}
\end{subequations}
By the definition of $\boper{D}$ [\Eq{eq:Ddef}] and by the properties of $\boper{\Pi}$ [\Eq{eq:Pipr}], this equation is invariant with respect to the gauge transformations \eq{eq:hpp0}.

\subsection{Geometrical optics of linear quasimonochromatic waves}
\label{sec:GO}

Let us consider a single quasimonochromatic wave
\begin{gather}
\tilde{h}^{\alpha\beta} = a^{\alpha\beta} \ee^{\ii\theta},
\quad 
\bar{k}_\mu \doteq \pd_\mu \theta,
\end{gather}
where the envelope $\smash{a^{\alpha\beta}}$ and the local wavevector $\bar{k}_\mu$ are slow compared with $\theta$ by a factor of $\epsilon \ll 1$. Using the Weyl expansion of the dispersion operator \cite{my:quasiop1, ref:mcdonald88}, \Eq{eq:Dh1} can be reduced to the following first-order equation:
\begin{gather}\label{eq:Dh12}
\vec{D}(x, \bar{k}) \vec{a}
- \ii \vec{V}^\alpha \pd_\alpha \vec{a}
- \frac{\ii}{2\sqrt{-g}}\,\frac{\pd (\sqrt{-g}\,\vec{V}^\alpha)}{\pd x^\alpha}\,\vec{a}= 0,
\end{gather}
where
\begin{gather}
\vec{V}^\alpha(x) \doteq \bigg[\frac{\pd \vec{D}(x, k)}{\pd k_\alpha}\bigg]_{k = \bar{k}(x)}.
\end{gather}
The field equation \eq{eq:Dh12} is gauge-invariant only in the limit $\epsilon \to 0$, but a strictly gauge-invariant theory can also be constructed at nonzero $\epsilon$, namely, as follows. 

As seen from \Eq{eq:Dh12}, $\smash{\boper{D} \tilde{\vec{h}}} \approx \smash{\vec{D}(x, \bar{k}) \tilde{\vec{h}}}$, so \Eq{eq:S2ql} can be reduced to
\begin{gather}\label{eq:aux11}
S^{(2)} = \frac{1}{4} \braket{\tilde{\vec{h}}| \boper{D} \tilde{\vec{h}}}
\approx \frac{1}{4} \braket{\tilde{\vec{h}}| \vec{D}(x, \pd\theta) \tilde{\vec{h}}}.
\end{gather}
At small $\epsilon$, one also has
\begin{gather}
\vec{D} \approx \vec{\Pi}^\dag \vec{\total{D}} \vec{\Pi} \equiv \vec{D}_0,
\end{gather}
where $\vec{\Pi}$ is the Weyl symbol of $\boper{\Pi}$ as in the Minkowski space \cite{my:gwinvar}:
\begin{gather}
\Pi^{\alpha\beta}{}_{\gamma\delta}
= \delta^\alpha_{(\gamma}\delta^\beta_{\delta)} 
- \frac{2k^{(\alpha}_{\phantom{\beta)}} \delta^{\beta)}_{(\gamma}k^{\phantom{\beta)}}_{\delta)}}{k^2}
+ g_{\gamma \delta}\,\frac{k^\alpha k^\beta}{k^2}.
\label{eq:PiMink}
\end{gather}
[Here, Greek indices are converted to Latin indices as in \Eq{eq:iot0}, $k^2 \doteq k_\mu k^\mu$, and the brackets in indices denote symmetrization, as usual.] Using this, let us approximate \Eq{eq:aux11}~as
\begin{gather}\label{eq:aux21}
S^{(2)}[\pd\theta, \vec{a}, \vec{a}^\dag] = \int \dd^4x\,\sqrt{-g}\,\mcc{L}(x, \pd\theta, \vec{a}, \vec{a}^\dag),
\end{gather}
where $\mcc{L}$ is the Lagrangian density given by
\begin{gather}\label{eq:mcL}
\mcc{L}(x, k, \vec{a}, \vec{a}^\dag) = \frac{1}{4}\,\vec{a}^\dag\vec{D}_0(x, k) \vec{a}
\end{gather}
and $\smash{\vec{a}^\dag}$ is a row vector with elements $\smash{a^{b*}}$. By the properties of $\vec{\Pi}$, this $\mcc{L}$ is exactly invariant with respect to the gauge transformation
\begin{gather}\label{eq:gi2}
a^{\alpha\beta} \to a'^{\alpha\beta} = a^{\alpha\beta} - 2\ii k^{(\alpha}\xi^{\beta)}
\end{gather}
(where $\xi^\alpha$ is an arbitrary vector field), which represents the GO limit of transformations \eq{eq:hpp0}. Hence, so is the whole field theory generated by \Eq{eq:aux21}. 

Because $\smash{S^{(2)}}$ has the standard form of the GO action that governs a linear dispersive wave, this theory is readily constructed as usual \cite{book:tracy, my:amc}.\footnote{Here, we consider only the lowest-order GO approximation, in which the gravitational spin-Hall effect \cite{ref:andersson21, my:quasiop1, my:covar} is neglected.} First, let us express the vector $\vec{h}$ through the scalar amplitude $a \doteq |\vec{a}|$ and the local polarization $\smash{\vec{e}} \doteq \smash{\vec{a}/|\vec{a}|}$. The polarization and the dispersion relation are found from \Eq{eq:Dh1}, which in the GO limit yields
\begin{gather}
\vec{D}_0(x, \bar{k})\vec{e} = 0,
\quad
\det \vec{D}_0(x, \bar{k}) = 0.
\end{gather}
Then, $\mcc{L}$ can be expressed as
\begin{gather}
\mcc{L}(x, k, a) = a^2 D(x, k), 
\quad
D \doteq \vec{e}{}^\dag\vec{D}_0 \,\vec{e}/4.
\end{gather}
Hence the dispersion relation can be expressed as 
\begin{gather}
D(x, \bar{k}) = 0.
\end{gather}
The corresponding flux density of the wave action is
\begin{gather}\label{eq:Jact}
\mc{J}^\alpha = -a^2 \bigg[\frac{\pd D(x, k)}{\pd k_\alpha}\bigg]_{k = \bar{k}(x)},
\end{gather}
or equivalently, $\mc{J}^\alpha = - a^2 \vec{e}^\dag \vec{V}_0^\alpha\vec{e}/4$, where
\begin{gather}
\vec{V}_0^\alpha(x) \doteq \bigg[\frac{\pd \vec{D}_0(x, k)}{\pd k_\alpha}\bigg]_{k = \bar{k}(x)}.
\end{gather}
The corresponding action-conservation theorem is \cite{my:amc}
\begin{gather}
\del_\alpha \mc{J}^\alpha = 0.
\label{eq:action}
\end{gather}
Similarly, the energy--momentum tensor of the wave is \cite{my:amc}
\begin{gather}\label{eq:Tab}
{\mc{T}_\alpha}^\beta = k_\alpha \mc{J}^\beta
\end{gather}
and satisfies
\begin{gather}
\del_\beta {\mc{T}_{\alpha}}^\beta = \pd_\alpha \mcc{L}.
\end{gather}
Note that $\smash{{\mc{T}_\alpha}^\beta}$ is the \textit{canonical} energy--momentum of a wave and, as such, does not have to be symmetric. On the definition of the wave energy--momentum, see, \eg \Refs{my:amc, ref:dewar77} and, particularly, \myRef{my:ql} for the relevance of the canonical energy--momentum in so-called oscillation-center QL theory. Together with the GO ray equations~\cite{book:tracy}
\begin{gather}
\frac{\dd x^\alpha}{\dd \tau} = \frac{\pd D(x, k)}{\pd k_\alpha},
\quad
\frac{\dd k_\alpha}{\dd \tau} = -\frac{\pd D(x, k)}{\pd x^\alpha},
\end{gather}
where $\tau$ is a parameter along the ray, the above equations provide a complete gauge-invariant GO theory of adiabatic GWs in a general medium.

\subsection{Example: quasimonochromatic vacuum GWs in the normal coordinates}
\label{sec:mink}

In the limit of vanishingly small density (and smooth enough background metric), one has $\boper{\total{D}} \approx \pd^2 \boper{\Pi} \approx \boper{\Pi}  \pd^2$, where $\pd^2 \doteq \pd_\mu \pd^\mu$ \cite{my:gwinvar}. Then, using \Eq{eq:Pipr}, one obtains
\begin{gather}
\boper{D} = \boper{\Pi}{}^\dag \pd^2 \boper{\Pi}{}^2 = \boper{\Pi}{}^\dag \pd^2 \boper{\Pi},\\
\vec{D}_0(x, k) = -\vec{\Pi}{}^\dag(x, k) k^2 \vec{\Pi}(x, k).
\end{gather}
Also, the GW dispersion relation in this case is $k^2 \to 0$, so $\vec{\Pi}$ becomes singular. However, $\smash{{\mc{T}_\alpha}^\beta}$ must remain finite, which means that $\vec{e}^\dag \vec{V}_0^\alpha\vec{e}$ must remain finite. 

Let us assume the normal coordinates, so that the metric is locally close to the Minkowski metric. Then, by performing a direct calculation, one finds that polarization in this case must satisfy
\begin{subequations}
\begin{gather}
e_{00} - 2 e_{03} + e_{33} = 0,\\
e_{01} - e_{13} = 0,\\
e_{02} - e_{23} = 0,
\end{gather}
\end{subequations}
at $\bar{k}^\alpha \to (\omega, 0, 0, \omega)$. One can recognize these as the conditions for certain metric invariants \cite{my:gwinvar} to be zero. One can also express $\mcc{L}$ in terms of metric invariants directly. Any combination of those are also invariants, and we choose to work with the following set (cf.\ \myRef{my:gwinvar}):
\begin{subequations}
\begin{gather}
\Psi_1 \equiv \Psi_+ = \frac{1}{2}\,(a_{11} - a_{22}),\\
\Psi_2 \equiv \Psi_\times = a_{12} = a_{21},\\
\Psi_3 = \frac{\sk}{\omega}\,a_{01} - a_{13},\\
\Psi_4 = \frac{\sk}{\omega}\,a_{02} - a_{23},
\end{gather}
and also
\begin{multline}
\Psi_{5,6} = \frac{2 \pm \sqrt{2}}{4}\,(a_{11} + a_{22}) (\omega^2 - \sk^2) \\
\pm \frac{1}{\sqrt{2}}\,(a_{00} - 2\omega k a_{03} + \omega^2 a_{33}),
\end{multline}
\end{subequations}
where we assumed
\begin{gather}
\bar{k}^\alpha = (\omega, 0, 0, \sk)
\end{gather}
with nonzero $\omega$. Then, a direct calculation shows that \Eq{eq:mcL} can be rewritten as follows:
\begin{multline}
\mcc{L} = 
\frac{\omega^2 - \sk^2}{2}\,(|\Psi_1|^2 + |\Psi_2|^2) 
\\
+ \frac{\omega^2}{2}\,(|\Psi_3|^2 + |\Psi_4|^2) 
+ \frac{|\Psi_5|^2 + |\Psi_6|^2}{2(\omega^2 - \sk^2)}.
\label{eq:L2}
\end{multline}
From 
\begin{gather}
\frac{\delta S^{(2)}}{\delta \Psi_a} = 0,
\quad
\frac{\delta S^{(2)}}{\delta \Psi_a^*} = 0,
\end{gather}
one finds that $\Psi_3 = \Psi_4 = \Psi_5 = \Psi_6 = 0$; in particular, $\Psi_5 = \Psi_6 = o(k^2)$ at $k^2 \to 0$. Assuming the notation $\vec{\Psi} \doteq (\Psi_+, \Psi_\times)^\intercal$, \Eq{eq:L2} can be simplified~as
\begin{gather}
\mcc{L} = -\frac{k^2}{2}\,(|\Psi_+|^2 + |\Psi_\times|^2) = -\frac{k^2}{2}\,|\vec{\Psi}|^2,
\label{eq:laggw}
\end{gather}
which describes the two well-known vacuum GW modes with dispersion relation $k^2 = 0$. Also, by \Eqs{eq:Jact} and \eq{eq:Tab}, the density of the wave action flux and the canonical energy--momentum tensor are given by
\begin{gather}\label{eq:Tvac}
\mc{J}^\alpha = k^\alpha |\vec{\Psi}|^2,
\quad
{\mc{T}_\alpha}^\beta = k_\alpha k^\beta |\vec{\Psi}|^2.
\end{gather}

\subsection{Quasimonochromatic vacuum GWs in general background coordinates}
\label{sec:gb}

The above results for the normal coordinates can be readily generalized to an arbitrary background coordinates. As usual \cite{book:whitham, my:itervar}, in order to obtain leading-order GO equations, one can approximate the wave Lagrangian density to the zeroth order in $\epsilon$.\footnote{Keeping \(\mc{O}(\epsilon^1)\) corrections only weakly affects the dispersion relation and introduces corrections to the action equation that are of order \(\epsilon^2\) and thus typically negligible on propagation distances of order of $\ell_g$. For a more formal justification of this approach, see, for example, \myRef{book:whitham}.} Then, the dispersion relation is still given by $k^2 = 0$ for both modes. By standard variational theory of linear waves \cite{book:whitham}, this immediately yields
\begin{gather}\label{eq:laggw2}
\mcc{L} = -\frac{k^2}{2}\,(A_+^2 + A_\times^2) = -\frac{k^2}{2}\,A^2.
\end{gather}
Here, $A_+$ and $A_\times$ are real functions quadratic in the wave amplitude (we define them as positive semi-definite for clarity), \(A \doteq (A_+^2 + A_\times^2)^{1/2}\), and the minus sign ensures that the wave energy be non-negative. Then, the corresponding action integral \eq{eq:aux21} becomes
\begin{gather}\label{eq:s2gb}
S^{(2)} = -\frac{1}{2} \int \dd^4x\,\sqrt{-g}\,k^2 A^2,
\end{gather}
where $\sqrt{-g}$ no longer can be replaced with unity.

Because $S^{(2)}$ and $k^2$ are scalars, and so is $d^4x\sqrt{-g}$, the functions $A_+$, $A_\times$, and $A$ are scalars. One can readily link them to the quantities that we have introduced above for the normal coordinates by comparing the Lagrangian densities \eq{eq:laggw} and \eq{eq:laggw2}:
\begin{gather}
A_+ = |\Psi_+|,
\quad
A_\times = |\Psi_\times|,
\quad
A = |\Psi|,
\end{gather}
and the elements of $a^{\alpha\beta}$ can be inferred from $A_+$ and $A_\times$ using that we already know the wave polarization (to the zeroth order in $\epsilon$, which is of interest here) in the normal coordinates. Finally, we find that
\begin{gather}\label{eq:JT2}
\mc{J}^\alpha = k^\alpha A^2,
\qquad
{\mc{T}_\alpha}^\beta = k_\alpha k^\beta A^2
\end{gather}
and can describe the evolution of the wave amplitude  using \Eqs{eq:action} and \eq{eq:Tab}. [The precise formulas for $\mc{J}^\alpha$ and ${\mc{T}_\alpha}^\beta$ in inhomogeneous medium contain small local corrections compared to \Eq{eq:JT2}, but those are bounded to remain  $\mc{O}(\epsilon)$ and vanish in the GO limit.] The former readily shows that $A$ respond nontrivially to the background inhomogeneity:
\begin{gather}
\frac{1}{\sqrt{-g}}\,\frac{\pd}{\pd x^\alpha}\,(\sqrt{-g}\,k^\alpha A) = 0,
\end{gather}
while \Eq{eq:Tab} becomes $\del_\beta {\mc{T}_\alpha}^\beta = 0$.

\section{Quasilinear theory}
\label{sec:QL}

\subsection{Basic equations}

Now let us consider the self-consistent evolution of $\smash{g^{\alpha\beta}}$ as an independent variable. Equations \eq{eq:Dh0} do not change in this case, but an additional equation emerges:
\begin{gather}
\frac{\delta S}{\delta g^{\alpha\beta}} = 0,
\label{eq:slieeq2}
\end{gather}
which can also be written as
\begin{gather}
G_{\alpha\beta} = T_{\alpha\beta} + \mc{N}_{\alpha\beta},\label{eq:Dg1}\\
\mc{N}_{\alpha\beta} \doteq - \frac{2}{\sqrt{-g}} \frac{\delta S^{(2)}}{\delta g^{\alpha\beta}},
\label{eq:Dg2}
\end{gather}
where \(G_{\alpha\beta}\) and \(T_{\alpha\beta}\) are the background Einstein tensor and the background energy-momentum tensor respectively, given by substituting the full metric with the background metric in \Eqs{eq:efe}. The equation above \eq{eq:Dg1} is invariant with respect to the gauge transformations \eq{eq:hpp0} by the invariance of $\smash{S^{(2)}}$. It is nonlinear in that $\mc{N}_{\alpha\beta}$ is quadratic in $\smash{h^{\alpha\beta}}$ and describes the slow evolution of the background metric in response to GWs. In contrast, direct interactions between GWs are neglected in this model, as seen from the linearity of \Eq{eq:Dh0}. We call this model QL by analogy with the QL models that are used in plasma physics to describe the interactions between plasmas and electromagnetic waves \cite{ref:dewar73, my:ql, my:qponder, my:wkeadv}. More specifically, it is an \textit{adiabatic} QL model in the sense that $\smash{g^{\alpha\beta}}$ and $\smash{h^{\alpha\beta}}$ are assumed to be the only independent variables in the system (see also \Sec{sec:metric}). 

\subsection{Alternative independent variables}
\label{sec:alt}

The expression for $\smash{S^{(2)}}$ in \Eq{eq:Dg2} can be taken from \Eq{eq:S2ql}. In this case, it may be more convenient to switch from independent variables $[g, \tilde{\vec{h}}, \tilde{\vec{h}}{}^*]$ to the independent variables $[g, \tilde{\vec{\mcc{h}}}, \tilde{\vec{\mcc{h}}}{}^*]$, where
\begin{gather}
\vec{\mcc{h}} \doteq (-g)^{1/4} \vec{h}, 
\quad
\vec{\mcc{h}} = \frac{1}{2}\,(\tilde{\vec{\mcc{h}}} + \tilde{\vec{\mcc{h}}}{}^*),
\end{gather}
where the second equality is introduced by analogy with \Eq{eq:comph}. Then, \Eq{eq:wigner} can be expressed as
\begin{multline}\label{eq:wigner2}
W^{ab}(x, k) \doteq \frac{1}{(2\pi)^4} \int \dd^4s\,\exp(-\ii k_\nu s^\nu)
\\
\times \favr{\mcc{h}^a(x + s/2)\,\mcc{h}^b(x - s/2)},
\end{multline}
so $\smash{\vec{W}}$ depends only on $\vec{\mcc{h}}$ but not on $g^{\alpha\beta}$. This shows that the functional derivative in \Eq{eq:Dg2} can be taken at fixed~$\smash{\vec{W}}$:
\begin{multline}
\mc{N}_{\alpha\beta} = - \frac{1}{\sqrt{-g}} \frac{\delta}{\delta g^{\alpha\beta}} 
\int \dd^4x\,\dd^4k \\
\times \vec{D}(x, k; [g]):{\vec{W}}(x, k; [\vec{\mcc{h}}, \vec{\mcc{h}}^*]),
\end{multline}
where $[g]$ and $[\vec{\mcc{h}}, \vec{\mcc{h}}^*]$ have been included in arguments to emphasize the dependence on the corresponding fields. Also, \Eq{eq:Dh} can be written as
\begin{gather}
\oper{\mcc{D}}_{\alpha\beta\gamma\delta}[g] \mcc{h}^{\gamma\delta} = 0,
\end{gather}
where we have introduced an alternative Hermitian dispersion operator
\begin{gather}
\boper{\mcc{D}} \doteq (-g)^{1/4} \boper{D} (-g)^{1/4}.
\end{gather}

\subsection{Example: vacuum GWs}
\label{sec:qlvac}

As an example, let us consider quasimonochromatic vacuum GWs in the general smooth background, which we previously discussed in \Sec{sec:gb}. Let us rewrite \Eq{eq:s2gb} as
\begin{gather}
S^{(2)}[g, \pd\theta, \bar{A}] = - \frac{1}{2}\int \dd^4 x\,g^{\alpha\beta} \bar{k}_\alpha \bar{k}_\beta \bar{A}^2,
\end{gather}
where $\bar{k}_\alpha \doteq \pd_\alpha\theta$ and we have also introduced $\bar{A} \doteq (-g)^{1/4}A$ as a new measure of the wave amplitude. By taking variation in \Eq{eq:Dg2} with respect to $\smash{g^{\alpha\beta}}$ at fixed $\theta$ and $A$, one obtains
\begin{gather}
\mc{N}_{\alpha\beta} = \bar{k}_\alpha \bar{k}_\beta A^2 = \mc{T}_{\alpha\beta},
\end{gather}
where we used \Eq{eq:Tvac}. Because $T_{\alpha\beta} = 0$ in vacuum, \Eq{eq:Dg1} becomes, expectedly,
\begin{gather}
G_{\alpha\beta} = \mc{T}_{\alpha\beta}.
\end{gather}

Similarly, for a continuous spectrum, we have
\begin{gather}
S^{(2)} = - \frac{1}{2}\int \dd^4 x\,\dd^2k\, g^{\alpha\beta} k_\alpha k_\beta {W}_\Psi.
\end{gather}
Here, $W_\Psi \doteq W_+ + W_\times$, $W_s$ are the Wigner functions of the two vacuum modes:
\begin{multline}
W_s(x, k) \doteq \frac{1}{(2\pi)^4} \int \dd^4s\,\msf{J}(x,s)\,\exp(-\ii k_\nu s^\nu)
\\
\times \Psi_s(x + s/2)\,\Psi_s^*(x - s/2),
\end{multline}
and $s = +,\times$. Hence, \Eq{eq:Dg1} can be expressed as
\begin{gather}\label{eq:Gv}
G_{\alpha\beta} = \frac{1}{\sqrt{-g}}\int \dd^4k\,k_\alpha k_\beta {W}_\Psi(x, k).
\end{gather}
Also note that the Wigner function is delta-shaped in this case, ${W}_\Psi\propto \delta(k^2)$, because the waves are constrained by the dispersion relation $k^2 = 0$. This means, in particular, that by taking the trace of \Eq{eq:Gv}, one finds that the corresponding Ricci scalar is zero.

\section{Gauge invariance beyond the quasilinear approximation}
\label{sec:euler}

There is also an alternative path to a gauge-invariant theory of GWs, which extends beyond the QL approximation and is formulated as follows. Let us represent the perturbation metric as $h_{\alpha\beta} = \hm \vartheta_{\alpha\beta}$, where $\vartheta_{\alpha\beta} = \mc{O}(1)$. Then, one can search for $\total{g}_{\alpha\beta}$ in the form
\begin{gather}
\total{g}_{\alpha\beta} = \sum_{n=0}^\infty \hm^n \vartheta^{(n)}_{\alpha\beta},
\end{gather}
with $\smash{\vartheta^{(n)}_{\alpha\beta} = \mc{O}(1)}$ and $a$ is now a constant that represents the characteristic GW amplitude. This implies
\begin{subequations}\label{eq:ghn}
\begin{gather}
g_{\alpha\beta} = \sum_{n=0}^\infty \hm^n \favr{\vartheta^{(n)}_{\alpha\beta}},
\label{eq:gn}
\\
h_{\alpha\beta} = \sum_{n=1}^\infty \hm^n 
\big[\vartheta^{(n)}_{\alpha\beta}-\favr{\vartheta^{(n)}_{\alpha\beta}}\big],
\label{eq:hn}
\end{gather}
\end{subequations}
because, by the previously stated assumption, \(h_{\alpha\beta} =\mc{O}(a)\). The variation $\sigma_{\alpha\beta}$ that enters the Einstein equations \eq{eq:Stot} can be represented as
\begin{gather}\label{eq:sigma}
\sigma_{\alpha\beta}[\total{g}] = \sum_{n=0}^\infty \hm^n\sigma_{\alpha\beta}^{(n)},
\quad
\sigma_{\alpha\beta}^{(n)} = \mc{O}(1).
\end{gather}
Then, using the Einstein equations $\sigma_{\alpha\beta} = 0$, one obtains an infinite set of equations for $\smash{\vartheta^{(n)}_{\alpha\beta}}$:
\begin{gather}\label{eq:s1}
\sigma_{\alpha\beta}^{(n)} = 0.
\end{gather}
These equations are invariant with respect to coordinate transformations \eq{eq:vt}, which is seen as follows. The functions $\smash{\sigma_{\alpha\beta}^{(n)}}$ are not true tensors but $\smash{\sigma_{\alpha\beta}}$ is, so $\sigma_{\alpha\beta} \to \smash{\sigma'_{\alpha\beta}}$ is a tensor transformation. For example, let us assume $\xi^\mu = \mc{O}(\hm)$; then \cite{book:carroll}, 
\begin{gather}\label{eq:spd}
\sigma_{\alpha\beta}' 
= \sigma_{\alpha\beta} - \lie_\xi \sigma_{\alpha\beta} + \mc{O}(\hm^2).
\end{gather}
Assuming a decomposition of $\smash{\sigma'_{\alpha\beta}}$ similar to \Eq{eq:sigma}, one obtains
\begin{subequations}\label{eq:sigp}
\begin{gather}
\sigma_{\alpha\beta}'^{(0)} = \sigma_{\alpha\beta}^{(0)},
\label{eq:sigp0}
\\
\sigma_{\alpha\beta}'^{(1)} = \sigma_{\alpha\beta}^{(1)} - \lie_\xi \sigma_{\alpha\beta}^{(0)},
\label{eq:sigp1}
\end{gather}
\end{subequations}
and similarly, $\smash{\sigma_{\alpha\beta}'^{(n)}}$ involves derivatives of $\smash{\sigma_{\alpha\beta}'^{(m)}}$ with $m < n$. Like in the original coordinates, the Einstein equations can be written as $\smash{\sigma_{\alpha\beta}' = 0}$, or equivalently, 
\begin{gather}\label{eq:s2}
\sigma_{\alpha\beta}'^{(n)} = 0.
\end{gather}
For $n = 0$, this yields $\smash{\sigma_{\alpha\beta}'^{(0)}}= 0$, which leads to $\smash{\sigma_{\alpha\beta}^{(0)}} = 0$ by \Eq{eq:sigp0}. Then, by \Eq{eq:sigp1}, one obtains $\smash{\sigma_{\alpha\beta}'^{(1)}} = \smash{\sigma_{\alpha\beta}^{(1)}}$, which leads to $\smash{\sigma_{\alpha\beta}^{(1)} = 0}$ by \Eq{eq:s2} with $n = 1$. By induction, \Eqs{eq:s2} are equivalent to \Eqs{eq:s1} also at $n > 1$. 

Similar arguments apply when $\xi^\mu$ scales with $\hm$ nonlinearly, say, as $\smash{\xi^\mu = \sum_n \hm^n \xi_{(n)}^\mu}$. In this case, \Eq{eq:spd} can be written (exactly) in the form
\begin{gather}\label{eq:spd2}
\sigma_{\alpha\beta}' 
= \sigma_{\alpha\beta} - \oper{L}(\xi) \sigma_{\alpha\beta},
\end{gather}
where $\oper{L}(\xi) \sigma_{\alpha\beta}$ denotes a term that is linear in $\sigma_{\alpha\beta}$ and, generally, nonlinear in $\smash{\xi^\mu}$. The  relations between $\smash{\sigma_{\alpha\beta}'^{(n)}}$ and $\smash{\sigma_{\alpha\beta}^{(n)}}$ depend on specific $\smash{\xi_{(n)}^\mu}$, but one can still apply the same arguments as in the previous case. Then, one again finds that \Eqs{eq:s2} are equivalent to \Eqs{eq:s1}.

Finally, notice the following. The functions \(\smash{\sigma^{(n)}_{\alpha\beta}}\) contain \(\smash{\vartheta^{(i)}_{\alpha\beta}}\) only with \(i \le n\). Hence, the set of equations \eq{eq:s1} can be truncated by considering only $0 \le n \le m$ with some finite $m$. In particular, note that the equation with $n = 1$, which governs \(\smash{\vartheta^{(1)}_{\alpha\beta}}\), is solved on the solution of \(\smash{\sigma^{(0)}_{\alpha\beta} = 0}\). This solution, \(\smash{\vartheta^{(0)}_{\alpha\beta}}\), is \textit{not} the background metric $g_{\alpha\beta}$. In order to find $g_{\alpha\beta}$ to the second order in $\hm$, one must solve the equation \(\smash{\sigma^{(2)}_{\alpha\beta}} = 0\) for \(\smash{\vartheta^{(2)}_{\alpha\beta}}\) and take the average as described in \Eq{eq:gn}. This makes our formulation different from that in \myRef{ref:isaacson68a}, where gauge invariance was lost because the second-order corrections were introduced directly into the zeroth-order equation.

\section{Summary}
\label{sec:sum}

In summary, we show how to keep the theory of dispersive GWs gauge-invariant beyond the linear approximation and, in particular, obtain an unambiguous gauge-invariant expression for the energy--momentum of a GW in dispersive medium [\Eq{eq:Tab}]. Using analytic tools from plasma physics, we propose an exactly gauge-invariant QL theory, in which GWs are governed by linear equations (\Sec{sec:linear}) but also affect the background metric on scales large compared to their wavelength (\Sec{sec:QL}). As a corollary, the gauge-invariant GO of linear dispersive GWs in a general background is formulated (\Sec{sec:GO}). As an example, we show how the well-known properties of vacuum GWs are naturally and concisely yielded by our theory in a manifestly gauge-invariant form (Secs.~\ref{sec:mink} and \ref{sec:qlvac}). We also show how the gauge invariance can be maintained within a given accuracy to an arbitrary order in the GW amplitude (\Sec{sec:euler}). These results are intended to form a physically meaningful framework for studying dispersive GWs in matter.

This material is based upon the work supported by National Science Foundation under the grant No. PHY~1903130.

%

\end{document}